\documentclass[12pt]{iopart}
\usepackage{iopams}
\usepackage{epsfig}
\input{epsf}
\begin{document}
\title[Flat Friedman universe  with
exponential potential scalar field and  dust]{Remarks on the general solution for the flat Friedman universe  with
exponential  scalar-field potential and  dust}
\author{A A Andrianov$^{1,2}$, F Cannata$^{3}$ and A Yu Kamenshchik$^{3,4,5}$}
\address{$^{1}$V.A. Fock Department of Theoretical Physics, Saint Petersburg State University, 198904, S.Petersburg, Russia\\
$^{2}$Departament d'Estructura i Constituents de la Materia and Institut de Ci\`encies del Cosmos (ICCUB)
Universitat de Barcelona, 08028, Barcelona, Spain\\
$^3$INFN, Sezione di Bologna, Via Irnerio 46, 40126 Bologna, Italy\\
$^{4}$Dipartimento di Fisica, Via Irnerio 46, 40126 Bologna,Italy\\
$^{5}$L.D. Landau Institute for Theoretical Physics of the Russian Academy of Sciences, Kosygin str.~2, 119334 Moscow, Russia}
\eads{\mailto{andrianov@bo.infn.it}, \mailto{cannata@bo.infn.it} and \mailto{kamenshchik@bo.infn.it}}
\begin{abstract}
We show that the simple extension of the method of obtaining  the general exact solution for the cosmological model with the
exponential scalar-field  potential to the case when the dust is present fails and we discuss the reasons of this puzzling
phenomenon.
\end{abstract}

\submitto{CQG}
\maketitle
\section{Introduction}
Exact solutions of the Einstein equations play a very important role in cosmology.  During the last decades the cosmological models
with scalar fields have acquired a great popularity. It is worth  mentioning various scenarios of the inflationary expansion of
the early universe \cite{inflation} or the quintessence models of the dark energy \cite{dark} responsible for the
phenomenon of cosmic acceleration \cite{cosmic}. However, the number of known exact solutions for cosmological models
based on scalar fields is rather limited. One of such models is the flat Friedman universe filled with a minimally coupled
scalar field with exponential potential. A particular solution for this model was known since the eighties and was studied in detail
\cite{power-law}--\cite{power-law5}. This solution describes a power-law expansion of the universe. More recently, the general solution of the Einstein equations
for this model was constructed \cite{general0}--\cite{general9}. This general solution was used for the description of such effects as transient acceleration,
and for the analysis of some models related to strings and branes. Besides, the solutions for the scalar fields with an exponential potential
and with a non-canonical kinetic terms were studied as well \cite{Chimento:2003ta}--\cite{Chimento:2010un}.
In papers \cite{Basilakos},\cite{Tsamparlis} the solutions for the flat Friedman universe with an exponential scalar-field  potential and dust were presented. We have made a direct check of these solutions and have found that they do not satisfy
the initial Friedman equations. The reason is that the authors of above mentioned papers using the method  of the time reparameterization  have made such a change in the Lagrangian and not in the equations of motion. The equations of motion
obtained from the modified Lagrangian describe a modified theory and not the original one. These two sets of equations of motion coincide in the case when the reparameterization invariance of the theory is explicit (see section 7) as it was in the case of the scalar field without dust. The addition of dust breaks this invariance. This issue seems to be rather  interesting  and instructive
from the methodological point of view and has stimulated us to write this short comment.
The structure of the paper is the following: in Sec. 2 we consider the model without dust and show that, generally the operations of time reparameterization and of the derivation of the equations of motion from the Lagrangian do not commute;
in Sec. 3 we illustrate this point considering a simple mechanical toy model; in Sec. 4 we come back to cosmology
and show that in the case of the Friedman model without dust the two sets of equations of motion coincide due to the
reparameterization invariance; the Sections 5 and 6 present explicit calculations for a specific value of the coupling
constant in the exponential potential in  both the models without and with dust while in the Sec. 7 we discuss what happens to
reparameterization invariance in the case with dust; the Section 8 is devoted to some concluding remarks.

\section{The model without dust}
Let us consider the possibility of the generalization of the general exact cosmological solution for the flat Friedman
universe filled with the spatially homogeneous scalar field with an exponential potential for the case when the dust-like
matter is present. The metric of the universe is
\begin{equation}
ds^2 = N^2(t)dt^2 - a^2(t)dl^2,
\label{Fried}
\end{equation}
where $N(t)$ is the lapse function and $a(t)$ is the cosmological radius.
Let us suppose that the universe is filled with the spatially homogeneous scalar field $\phi(t)$ with the potential $V(\phi)$.
One can show that the Hilbert-Einstein action for such a system can be reduced to the following Lagrangian, depending on $N(t), a(t)$
and $\phi(t)$ (here we have chosen a convenient normalization of the Newton gravitational constant):
\begin{equation}
L = \frac{\dot{a}^2a}{N} - \frac{a^3\dot{\phi}^2}{2N}+Na^3V(\phi),
\label{Lagr}
\end{equation}
where ``dot''
signifies the derivatives with respect to the cosmic time parameter $t$.
Making variation of this Lagrangian with respect to the lapse function $N$ and putting after that the value of $N$ equal to 1, we obtain the following first Friedman equation:
\begin{equation}
h^2 = \frac{\dot{\phi}^2}{2} + V,
\label{Fried1}
\end{equation}
where the Hubble parameter $h$ is
\begin{equation}
h \equiv \frac{\dot{a}}{a}.
\label{Hubble}
\end{equation}
The variation with respect to $a$ gives the second Friedman equation
\begin{equation}
\dot{h} + \frac{3}{2}h^2+\frac{3}{4}\dot{\phi}^2-3V = 0
\label{Fried2}
\end{equation}
and the variation with respect to the scalar field gives the Klein-Gordon equation
\begin{equation}
\ddot{\phi} +3h\dot{\phi} + \frac{dV}{d\phi} = 0.
\label{KG}
\end{equation}
The system of equations (\ref{Fried1}), (\ref{Fried2}) and (\ref{KG}) is not independent. One can consider the equations (\ref{Fried2}) and
(\ref{KG}) as a system of two equations for two variables ($a$ and $\phi$) and Eq. (\ref{Fried1}) is a first integral (or, in other words,
 a conservation law) of this system.

Now let us consider a particular potential:
\begin{equation}
V(\phi) = V_0\exp(\lambda \phi).
\label{poten}
\end{equation}
Let us introduce a pair of new variables $u$ and $v$ such that:
\begin{equation}
\phi = \frac{\sqrt{2}}{3}(v-u)
\label{vu}
\end{equation}
and
\begin{equation}
a^3 = e^{v+u}.
\label{vu1}
\end{equation}
Substituting these new variables into the Langrangian (\ref{Lagr}) (where $N$ is chosen to be equal to 1) we obtain
\begin{equation}
L = e^{v+u}\left(\frac43\dot{v}\dot{u} + V_0e^{\frac{\sqrt{2}\lambda}{3}(v-u)}\right).
\label{Lagr1}
\end{equation}
Making the variation of this Lagrangian with respect to the new variables $v$ and $u$, we obtain the following couple of the Euler-Lagrange
equations
\begin{equation}
\ddot{u}+\dot{u}^2 -  \left(\frac94+\frac{3\sqrt{2}\lambda}{4}\right)V_0e^{\frac{\sqrt{2}\lambda}{3}(v-u)} = 0,
\label{u}
\end{equation}
\begin{equation}
\ddot{v}+\dot{v}^2 +  \left(\frac{3\sqrt{2}\lambda}{4}-\frac94\right)V_0e^{\frac{\sqrt{2}\lambda}{3}(v-u)} = 0.
\label{v}
\end{equation}
These equations naturally coincide with those which can be obtained from Eqs. (\ref{Fried2}) and (\ref{KG}) by the substitution of the formulae
(\ref{vu}) and (\ref{vu1}).

Now let us introduce a new time parameter $\tau = \tau(t)$. In this case
\begin{eqnarray}
&&\dot{v} = v'\dot{\tau},\nonumber \\
&&\ddot{v} = v''\dot{\tau}^2+v'\ddot{\tau},\nonumber \\
&&\dot{u} = u'\dot{\tau},\nonumber \\
&&\ddot{u} = u''\dot{\tau}^2+u'\ddot{\tau},
\label{tau}
\end{eqnarray}
where ``prime'' denotes the derivative with respect to $\tau$.
It is convenient to choose the new time parameter $\tau$ in such a way that
\begin{equation}
\dot{\tau} = \frac32 \sqrt{V_0}e^{\frac{\sqrt{2}\lambda(v-u)}{6}}.
\label{tau1}
\end{equation}
Now, substituting the formulae (\ref{tau}) and (\ref{tau1}) into Eqs. (\ref{u}) and (\ref{v}) we obtain the following system of equations:
\begin{equation}
u'' +\left(1-\frac{\sqrt{2}\lambda}{6}\right)u'^2 +\frac{\sqrt{2}\lambda}{6}u'v'-\left(\frac{\sqrt{2}\lambda}{3}+1\right) = 0,
\label{u1}
\end{equation}
\begin{equation}
v'' +\left(1+\frac{\sqrt{2}\lambda}{6}\right)v'^2 -\frac{\sqrt{2}\lambda}{6}u'v'+\left(\frac{\sqrt{2}\lambda}{3}-1\right) = 0.
\label{v1}
\end{equation}
Obviously, these two equations are not disentangled due to the presence of the terms, proportional to $u'v'$.

Let us change the strategy and  in the action, corresponding to the Lagrangian (\ref{Lagr1}), make a transition to another time variable $\tau$. The new Lagrangian has the form
\begin{equation}
L_{new} = (v'u'+1)e^{(v+u)+\frac{\sqrt{2}\lambda}{6}(v-u)}.
\label{Lagr-new}
\end{equation}
The Euler-Lagrange equations, obtained from the Lagrangian (\ref{Lagr-new}) by variation with respect to $v$ and $u$ are
\begin{equation}
u'' +\left(1-\frac{\sqrt{2}\lambda}{6}\right)u'^2 -\left(\frac{\sqrt{2}\lambda}{6}+1\right) = 0,
\label{u2}
\end{equation}
\begin{equation}
v'' +\left(1+\frac{\sqrt{2}\lambda}{6}\right)v'^2 +\left(\frac{\sqrt{2}\lambda}{6}-1\right) = 0.
\label{v2}
\end{equation}
These two equations are disentangled and differ from Eqs. (\ref{u1}) and (\ref{v1}).
It is not strange because the theories, described by the Lagrangians (\ref{Lagr1}) and (\ref{Lagr-new}) are different.
The change of the time variable changes the theory. In other words, the operations of the change of time variables and the variation of the
action with respect to dynamical variables do not commute. To show this we shall consider a very simple toy model example in the following section.

\section{A toy model and modification of Euler-Lagrange equations}
Let us consider first a one-dimensional free non-relativistic particle. The only dynamical variable is the coordinate $x$ depending on the time $t$.
The Lagrangian of the model is
\begin{equation}
L = \frac{\dot{x}^2}{2},
\label{Lagr-toy}
\end{equation}
the equation of motion is
\begin{equation}
\ddot{x} = 0,
\label{eq-toy}
\end{equation}
and its general solution is
\begin{equation}
x(t) = x_0+vt,
\label{sol-toy}
\end{equation}
where $x_0$ and $v$ are arbitrary constants.

Let us introduce a new time parameter $\tau$ such that
\begin{equation}
\dot{\tau} = x.
\label{tau-toy}
\end{equation}
The new Lagrangian becomes
\begin{equation}
L_{new} = \frac{x'^2x}{2}.
\label{Lagr-toy1}
\end{equation}
The equation of motion is
\begin{equation}
x''x + \frac{x'^2}{2} = 0.
\label{eq-toy1}
\end{equation}
Its general solution is
\begin{equation}
x(\tau) = C(\tau + \tau_0)^{2/3},
\label{sol-toy1}
\end{equation}
where $C$ and $\tau_0$ are arbitrary constants.

Now, let us try to come back to the old time variable $t$. Substituting the formula (\ref{sol-toy1}) into Eq. (\ref{tau-toy}) we
obtain the equation
\begin{equation}
\frac{d\tau}{dt} = C(\tau+\tau_0)^{2/3},
\label{tau-toy1}
\end{equation}
whose general solution is
\begin{equation}
(\tau+\tau_0) = \left(\frac{C(t+t_0)}{3}\right)^3.
\label{sol-3}
\end{equation}
Substituting this formula into Eq. (\ref{sol-toy1}) we obtain
\begin{equation}
x(t) = \frac{C^3}{9}(t+t_0)^2,
\label{sol-x}
\end{equation}
which  has the functional dependence on $t$ which differs from (\ref{sol-toy}).

In a more general setting we proceed to the Lagrangian $L(\dot x, x)$ and perform the time reparameterization,
\[ \frac{d\tau}{dt} = f(x, t) .\]
The corresponding change in the action reads,
\begin{equation}
\int dt L(\dot x, x) \Longrightarrow \int d\tau \frac{1}{f} L(x'\cdot f, x),
\end{equation}
which gives the modified Euler-Lagrange equations,
\begin{equation}
\left(\frac{\partial L}{\partial x} - f d_\tau \Big(\frac{\partial L}{\partial (\dot x)}\Big)\right) + \frac{f'}{f}\left(f x' \frac{\partial L}{\partial (\dot x)} -L\right) = 0.
\end{equation}
When going back to the initial time variable one finds that vanishing of the first term in parentheses
 leads to the Euler-Lagrange equation of the original theory whereas the second term accumulates the deviation from it and its vanishing guarantees the equivalence of modified and original theories. Thus the two theories give the same classical dynamics if:
a) time reparameterization does not depend on dynamical variables, $f' = 0$; or b) reparameterization invariance holds, $\dot x \frac{\partial L}{\partial (\dot x)} -L = 0$ .
The latter case needs introduction of the lapse function to provide it.
\section{Back to the cosmology}
Now, one can notice that if we solve disentangled couple of equations (\ref{u2}) and (\ref{v2}), then, choosing some initial conditions,
the solutions will satisfy the equations of the initial theory with the Lagrangian (\ref{Lagr}). Why does this happen ?
The point is that the Lagrangian (\ref{Lagr1}) does not describe the initial cosmological model.
Indeed, in the initial Lagrangian (\ref{Lagr}) there is also another variable - the lapse function $N$. The variation with respect to this variable
gives us the first Friedman equation (\ref{Fried1}). Now, if we rewrite this equation in terms of the new variables $u$ and $v$ and the new time parameter $\tau$, we obtain
\begin{equation}
u'v' = 1.
\label{new-Fried1}
\end{equation}
Substituting the relation (\ref{new-Fried1}) into Eqs. (\ref{u1}) and (\ref{v1}) we see that these equations coincide with the disentangled equations
(\ref{u2}) and (\ref{v2}). Hence, solving Eqs. (\ref{u2}) and (\ref{v2}) and choosing the constants in the general solutions in such a way that the
relation (\ref{new-Fried1}) is satisfied we find the solutions which satisfy the original equations (\ref{u1}) and (\ref{v1}) and to the Friedman and
Klein-Gordon equations of the original theory. We shall illustrate it in the next section, considering the simplest case when  the coupling
constant $\lambda = \sqrt{18}$.
\section{The case $\lambda = \sqrt{18}$ without dust}
In the case when $\lambda = \sqrt{18}$, Eqs. (\ref{u2}) and (\ref{v2}) have the form
\begin{equation}
u'' = 2,
\label{u3}
\end{equation}
\begin{equation}
v''+2v'^2 = 0.
\label{v3}
\end{equation}
Their general solutions are
\begin{equation}
u = \tau^2 + u_1\tau + u_0,
\label{u-sol}
\end{equation}
\begin{equation}
v = v_0 + \frac12\ln (\tau + v_1),
\label{v-sol}
\end{equation}
where $u_0,u_1,v_0,v_1$ are arbitrary constants.
Now, substituting these general solutions into the expressions for $h,\phi$ and $\dot{\phi}$ we find
\begin{equation}
h = \frac{1}{2}\sqrt{V_0}e^{v-u}\left(\frac{1}{2(\tau+v_1)}+2\tau + u_1\right),
\label{Hubble0}
\end{equation}
\begin{equation}
h^2 = \frac{1}{4}V_0e^{2(v-u)}\left(\frac{1}{2(\tau+v_1)}+2\tau + u_1\right)^2,
\label{Hubble1}
\end{equation}
\begin{equation}
\dot{\phi} = \frac{\sqrt{2}}{2}\sqrt{V_0}e^{v-u}\left(\frac{1}{2(\tau+v_1)}-2\tau - u_1\right),
\label{phi-dot}
\end{equation}
\begin{equation}
\dot{\phi}^2 = \frac{1}{2}V_0e^{2(v-u)}\left(\frac{1}{2(\tau+v_1)}-2\tau - u_1\right)^2.
\label{phi-dot1}
\end{equation}
Substituting these expressions into the Friedman equation we find
\begin{equation}
h^2 - \frac12\dot{\phi}^2 - V(\phi) = V_0e^{2(v-u)}\left(\frac{u_1-2v_1}{2(\tau+v_1)}\right).
\label{new-Fried2}
\end{equation}
The right-hand side of Eq. (\ref{new-Fried2}) is equal to zero, and, hence, the first Friedman equation is
satisfied if and only if
\begin{equation}
u_1 = 2v_1.
\label{condition}
\end{equation}
It is easy to see that the condition (\ref{condition}) coincides with the condition (\ref{new-Fried1}) for the solutions (\ref{u-sol})--(\ref{v-sol}).

\section{The case $\lambda = \sqrt{18}$ with dust}
In the case when the dust is present in the universe the first Friedman equation has the following form
\begin{equation}
h^2 - \frac12\dot{\phi}^2 - V(\phi) = \frac{\rho_0}{a^3},
\label{new-Fried-dust}
\end{equation}
where $\rho_0$ is a constant, characterizing the quantity of the dust-like matter in the universe.
That means that from Eq. (\ref{new-Fried2}) and Eq. (\ref{new-Fried-dust}) it follows
\begin{equation}
 V_0e^{2(v-u)}\left(\frac{u_1-2v_1}{2(\tau+v_1)}\right) = \rho_0e^{-(v+u)},
 \label{cond-dust}
 \end{equation}
 or, in other terms,
 \begin{equation}
 V_0e^{3v-u)}\left(\frac{u_1-2v_1}{2(\tau+v_1)}\right) = \rho_0.
 \label{cond-dust1}
 \end{equation}
Substituting into Eq. (\ref{cond-dust1}) the explicit expressions for $u$ and $v$ from Eqs. (\ref{u-sol}) and (\ref{v-sol})
we obtain
\begin{equation}
\frac{V_0\sqrt{\tau+v_1}}{2}e^{3v_0-\tau^2-u_1\tau-u_0} = \rho_0.
\label{cond-dust2}
\end{equation}
It is easy to see that the condition (\ref{cond-dust2}) cannot be satisfied by a proper choice of the constants.
Thus, the solutions (\ref{u-sol}) and (\ref{v-sol}), which basically reproduce (up to a  change of notations) those, presented in the section IV.B. of the
paper \cite{Basilakos} do not satisfy the first Friedman equation and the solution presented in \cite{Basilakos}) for the flat Friedman universe
filled with the scalar field with the exponential potential and the dust is wrong. We shall try to explain the origin of the discrepancy in the following section.

\section{Reparameterization invariance and dust}
It is not strange that the solution of the equations of motion (\ref{u2}) and (\ref{v2}) following from the Lagrangian (\ref{Lagr-new}) obtained by the reparameterization of the time parameter does not satisfy the original first Friedman equation with dust. As it was emphasized before, the reparameterization of time, carried out not at the level of equations of motion but at the level of Lagrangian (action) gives a different theory
(see a simple toy model example in section 3). The question, however, arises: why in the case without dust one can obtain the correct solution, starting from the couple of equations  of motion  (\ref{u2}) and (\ref{v2}), derived from the modified Lagrangian ? The answer is based on the fact that the reparameterization of time
does not change the Lagrangian of the theory if this theory possesses the invariance with respect to reparameterization.
As is well known
the gravity theory
belongs to the class of the theories invariant with respect to reparameterization (see e.g. \cite{book}). However, both the Lagrangians (\ref{Lagr1}) and (\ref{Lagr-new}) are
not invariant with respect to the reparameterization, because in  both these Lagrangians the specific gauge is chosen - namely the lapse function $N$ is chosen to be equal to 1. Nevertheless, we can restore this gauge (reparameterization) invariance adding to the set of equations of motion the
first Friedman equation which is nothing but the constraint, corresponding to this invariance. In our case this constraint amounts
to the condition (\ref{condition}) which makes the system of disentangled equations (\ref{u2}) and (\ref{v2}) identical to the original system of equations (\ref{u1}) and (\ref{v1}).

Why this does not work in the presence of dust ? In the presence of dust the constraint, providing the reparameterization invariance of the
theory (i.e. the first Friedman equation) includes the dust term, proportional to $e^{v+u}$ that means that the product $u'v'$ is not equal
to 1, but represents some function of $u$ and $v$ and its substitution into the original couple of equations (\ref{u1}) and (\ref{v1})
does not transform  into the disentangled couple of equations (\ref{u2}) and (\ref{v2}).


\section{Concluding remarks}
We have seen that the simple extension of the method of obtaining  the general solution for the cosmological model with the
exponential potential scalar field to the case when the dust is present fails and we have tried to explain the reasons.
In paper \cite{Tsamparlis} the more general model was considered, where not only the scalar field and dust, but also a third
barotropic fluid was present. To disentangle the corresponding equations of motion the authors of \cite{Tsamparlis} have
used not only the change of the variables and the time reparameterization of the Lagrangian, similar to that used in
papers, studying the models with the exponential potential, but also an additional change of variables,
reducing these equations to the equations of the Ermakov-Pinney type \cite{Ermakov}. We have tried to use this second change
of variables in the model with the scalar field and dust at the level of the equations of motion (and not at the level
of the Lagrangian). We have seen, that in such a case the decoupling of the correct equations of motion does not occur.
Thus, the question about the generalization of the solution of the cosmological equations in the model with an exponential potential  scalar field for the case with dust, remains open.

\ack{The work of A.K. was partially supported by the RFBR grant No 11-02-00643. The work of A.A. is supported by SPbSU grant 11.0.64.2010 and also by projects FPA2010-20807, 2009SGR502, CPAN (Consolider CSD2007-00042).}

\end{document}